\documentclass{article}
\usepackage{graphicx}
\usepackage{latexsym}

\begin{document}
%\begin{flushright}
%Some fun numbers\\
%go here
%\end{flushright}
\begin{center}
{\huge 	Weak Scale Quantum Gravity Effects for $\gamma\gamma\to ZZ$ in
the TeV region.}\\
\hspace{10pt}\\
S.R.Choudhury\footnote{src@ducos.ernet.in} \\
{\em Department of Physics, Delhi University, Delhi 110007, India},\\
A.S.Cornell\footnote{a.cornell@tauon.ph.unimelb.edu.au}
and G.C.Joshi\footnote{joshi@physics.unimelb.edu.au}\\
{\em School of Physics, University of Melbourne,}\\
{\em Parkville, Victoria 3010, Australia}\\
\hspace{10pt}\\
$28^{th}$ of February, 2002
\end{center}
\hspace{10pt}\\
\begin{abstract}
\indent We consider the effect of massive spin-2 gravitons that occur as
Kaluza-Klein excitations of the graviton in a Weak scale Quantum Gravity
scenario on the process $\gamma \gamma \to ZZ$,  which in the Standard
Model can proceed through loop diagrams.  For a wide range of parameters,
we show that the massive gravitons leave behind signatures that should be
verifiable in a TeV scale scattering experiment.
\end{abstract}

\section{Introduction}
\hspace{10pt} Arkani-Hamed, Dimopoulos and Dvali \cite{one} have proposed
a higher $(4+n)$-dimensional spacetime theory in which the Standard 
Model (SM) particles live in the usual 4-dimensional spacetime where as
gravity also propagates in the additional compactified spatial 
dimensions.  A remarkable feature of this scenario is the
possibility that the Planck mass can become comparable to the electro-weak
scale, thus solving the hierarchy mystery that naturally exists in 
conventional theories where the Planck mass is many orders of magnitude
higher than the electro-weak scale. Gravity in our (3+1)-dimensional world
then becomes feeble for distances much larger than the compactification
scale $R$.  If $M_{Pl}$ is the Planck scale in our world and $M$ in the
$(4+n)$-dimensional world, the two are related by \cite{one} 
\begin{eqnarray}
M_{Pl}^2=R^nM^{2+n}
\end{eqnarray}
Setting $M\sim$ TeV, gives $R\sim$eV-MeV for $n=2-7$.  Noting that the 
Newtonian inverse square law of gravity has been verified up to about
a millimetre, \cite{two} we see that no contradiction with Newtonian
gravity results
in this approach.\\
\indent	An alternative scenario has been  proposed by Randall and Sundrum
\cite{three} where once again there exists only one fundamental Planck
Scale
($\sim$ TeV) with the hierarchy of scales been generated by an exponential
factor of the compactification radius in a 5-dimensional non factorizable 
geometry with the fifth dimension in the form of a torus.  Two 3-branes
with opposite tensions are assumed to reside at $S_1/Z_2$ orbifold fixed 
points $\phi=0,\pi$ where $\phi$ is the angular coordinate parametrizing
the extra dimension.  Of the two 3-branes, it is assumed that the
SM-fields reside only at the one at $\phi=\pi$ whereas gravity is all
over.  Calling as before $M_{Pl}$ the 4-dimensional Planck scale, $M$ the
Planck scale in the 5-dimensional world, $r_c$ the compactification
radius of the 5-dimensional torus, the two Planck scales are related by
the equation
\begin{eqnarray}
M_{pl}^2=\frac{M^3}{k}(1-e^{-2kr_c\pi}) ,
\end{eqnarray}
where $k \sim M_{Pl}$ is a parameter of the theory.  Choosing 
$kr_c\sim 12$, one sees that $M$ can have values in the TeV range, thus
providing an alternative solution to the hierarchy problem.\\
\indent	The graviton propagating in the compact dimensions, in addition to
our 4-dimensional world, appear in our world as a massless particle
together with its radial excitations.  In the ADD-scenario, these set of
resonances are almost a continuum, each individually coupling with the
4-dimensional Planck scale strength.  However because of the large number
of these resonances, their combined effect mimic a weak scale spin-2
excitation.  In the RS model, the graviton excitations are discrete with
each coupling with masses calculable in terms of k and $r_c$ and each
coupling with TeV scale strength with matter.\\
\indent Turning to phenomenological consequences of WSQG, it is clear that
there will be in the nature of effective exchange effects of spin-2
particles with TeV scale coupling.  Typically such contributions to the
amplitudes will a have suppression factor of $\left( \sqrt{s}/M_s
\right)^{n+2}$ replacing the coupling parameters $e$ or $g$ in the
corresponding SM amplitude.  Clearly the most important processes where
these effects will show up in the sub-TeV range are the ones where the
SM-contribution is suppressed occurring only through higher order loop
diagrams.  Processes involving $\gamma\gamma$ collisions like
$\gamma\gamma\to ZZ$ offer the most exciting possibility of such
processes.  The SM contributions through loops of all charged particles
are well calculated and at the same time the process seems to be
experimentally accessible with $\gamma-$rays obtained from backward 
scattering of laser beams off high energy electrons.  In the present
investigation, we estimate the signatures of the WSQG theories in this
process for energies in the TeV-range.

\section{SM-Amplitudes}
\hspace{10pt}The process
\begin{eqnarray}\label{react}
\gamma(p_1,\lambda_1)+\gamma(p_2,\lambda_2) \rightarrow 
Z(p_3,\lambda_3)+Z(p_4,\lambda_4)
\end{eqnarray}
is described by 36 invariant amplitudes $F(s,t,u ;\lambda_1, \lambda_2,
\lambda_3, \lambda_4)$ where the quantities $p$, $\lambda$ in parentheses
in equation (\ref{react}) denote the momentum and helicity of the 
corresponding particles and $s,t,u$ are the Mandelstam variables.  The
$F$'s are related to the corresponding differential cross-section by the
relation
\begin{eqnarray}
\frac{d\sigma}{d\cos\theta}=\frac{\beta_z}{64\pi s} |F|^2
\end{eqnarray}
where $\theta$ is the c.m. scattering angle and $\beta_z^2 =
(1-\frac{4m_z^2}{s})$.  Parity and Bose statistics imposes the following
restrictions on the amplitudes
\begin{eqnarray}
F(s,t,u; \lambda_1\lambda_2\lambda_3\lambda_4) & = & 
F(s,t,u; \lambda_2\lambda_1\lambda_4\lambda_3)(-1)^{\lambda_3-\lambda_4}\\
F(s,t,u; \lambda_1\lambda_2\lambda_3\lambda_4) & = & 
F(s,u,u;\lambda_2\lambda_1\lambda_3\lambda_4)(-1)^{\lambda_3-\lambda_4}\\
F(s,t,u; \lambda_1\lambda_2\lambda_3\lambda_4) & = &
F(s,u,t; \lambda_1\lambda_2\lambda_4\lambda_3) 
\end{eqnarray}
and
\begin{eqnarray}
F(s,t,u; \lambda_1\lambda_2\lambda_3\lambda_4) = 
F(s,t,u;-\lambda_1,-\lambda_2,-\lambda_3,-\lambda_4) .
\end{eqnarray}
The SM amplitudes corresponding to the W-loop and the fermion loops 
(figure 1a) have been evaluated \cite{four}.  In the high energy limit
that we are interested in, namely  $s,-t,-u \gg M_{Pl}^2$ most of the
amplitudes become negligible.  For the W-loop contribution, the surviving
amplitudes are the following together of course with their symmetry
counterparts;
\begin{eqnarray}
F^W (s,t,u; \lambda_1,\lambda_2,\lambda_3,\lambda_3) &=& 
\frac{\alpha^2}{s_w^2} 
A^w(s,t,u; \lambda_1,\lambda_2,\lambda_3,\lambda_4) \nonumber \\
A^w (s,t,u; \lambda_1,\lambda_2,\lambda_3,\lambda_4) &=& 
\frac{(12c_w^4-4c_w^2+1)}{4 c_w^2} 
A^s(s,t,u; \lambda_1,\lambda_2,\lambda_3,\lambda_4) \nonumber \\
&& \hspace{6em} + 
\delta^w(s,t,u; \lambda_1,\lambda_2,\lambda_3,\lambda_4) \nonumber \\
A^s (s,t,u; ++++) &=& 4 - \frac{4ut}{s^2}\left[ \log^2
\left| \frac{t}{u} \right|+\pi^2 \right]+4\frac{(t-u)}{s} \log
\left| \frac{t}{u} \right| \nonumber \\
A^s (s,t,u; +-+-) &=& 4 - \frac{4st}{u^2} \left[ \log^2 \left|
\frac{s}{t} \right| - 2i\pi \log \left| \frac{s}{t} \right| \right]
\nonumber \\
&& \hspace{4em} + 4 \frac{(s-t)}{u} \left[ \log \left| \frac{s}{t}
\right|- i \pi \right] \nonumber \\
\delta^w(s,t,u; ++++) &=& 16 c_w^2  \left\{ log^2 \left| \frac{t}{u}
\right| + \pi^2 + \frac{s}{u} \log \left| \frac{u}{m_w^2}\right| \log
\left| \frac{s}{t} \right| \right. \nonumber \\
&& \hspace{4em} + \frac{s}{t} \log \left| \frac{t}{m_w^2}\right| \log
\left|\frac{s}{u}\right| \nonumber \\
&& \hspace{5em} - \left. i \pi \left[\frac{s}{u} \log
\left| \frac{u}{m_w^2} \right| + \frac{s}{t} \log \left| \frac{t}{m_w^2}
\right| \right] \right\} \nonumber \\
\delta^w(s,t,u; +-+-) &=& 16 c_w^2 \left\{ \log^2 \left| \frac{t}{s}
\right| + \frac{u}{t} \log \left| \frac{t}{m_w^2} \right|
\log \left| \frac{u}{s} \right| \right. \nonumber \\
&& \hspace{4em} + \frac{u}{s} \log \left| \frac{s}{m_w^2} \right| \log
\left| \frac{u}{t} \right| + i \pi \left[ \frac{u}{t} \log
\left|\frac{s}{M_w^2} \right| \right. \nonumber\\
&& \hspace{6em} \left. \left. - \frac{u}{s} \log \left| \frac{u}{s}
\right| -  \frac{s^2+t^2}{st} \log \left| \frac{t}{s} \right|
\right] \right\} 
\end{eqnarray}
For the fermion loops, similarly only a few amplitudes (and their 
symmetric counterparts) survive as follows:
\begin{eqnarray}
F^f\left( s,t,u; \lambda_1,\lambda_2,\lambda_3,\lambda_4 \right) &=&
\frac{\alpha^2 Q_f^2}{4 s_w^2 c_w^2} \left\{ \left[ t_3^f-2 Q_f s_w^2
\right]^2 A^{vf}_{\lambda_1, \lambda_2, \lambda_3, \lambda_4}
(s,t,u,m_f) \right. \nonumber \\
&& \hspace{4em} + \left. (t^f_3)^2 
A^{af}_{\lambda_1,\lambda_2,\lambda_3,\lambda_4} \right\}  \nonumber \\
A^{vf} &=& -2 A^s + \delta^{vf}\nonumber \\  
\mathrm{and} \hspace{2em} A^{af} &=& -2 A^s + \delta^{af}  
\end{eqnarray}
with the following nonvanishing $\delta$'s  for both the superscripts:

\begin{eqnarray}
\delta_{++++}&=& -4 \left[ \log^2 \left| \frac{u}{t} \right| + \pi^2
\right] \nonumber \\
\delta_{+-+-}&=& -4 \left[ \log^2 \left| \frac{s}{t} \right| - i 2\pi
\log \left| \frac{s}{t} \right| \right]
\end{eqnarray}

\begin{figure}
\includegraphics[angle=0,width=10cm]{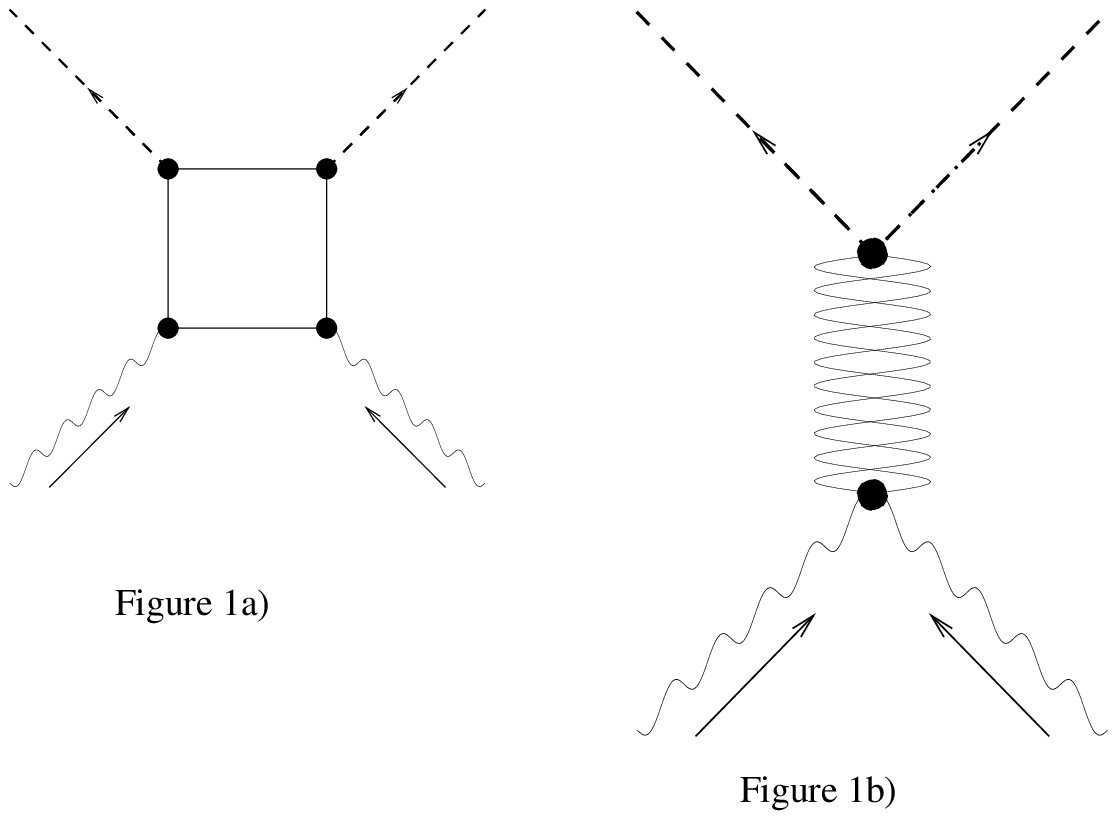}
\caption{a)  the SM loop graph - one single graph for all charged
particles in the loop, where the dashed lines represent the
$Z$ particles.\hspace{12em}  
Figure 1: b)  The $s$-channel graviton contribution graph,
where the double sinusoid represents the graviton.}
\label{fig1}
\end{figure}
	
\section{The Massive Graviton Contribution}
\hspace{10pt} Spin-2 massive excitations of the graviton contribute to
the amplitudes via $s$-channel pole, as shown in fig.1(b).  In the ADD 
version of the WSQG, the interaction of spin-2 excitations 
$h^{(n)}_{\mu\nu}$ of mass $m$ is given by the interaction Lagrangian
\begin{eqnarray}
{\cal L}_I = - \frac{\kappa}{2} \sum_n \int d^4x T^{\mu\nu}
h^{(n)}_{\mu\nu} ,
\end{eqnarray}
where $\kappa = \sqrt{16\pi G_N}$ and $T^{\mu\nu}$ is the energy momentum
tensor.  Following standard Feynman rules, the contribution of
fig.1(b) to the amplitudes are given by;
\begin{eqnarray}
F^g(s,t,u; + + \lambda_3 \lambda_4) = 0 \nonumber \\
\end{eqnarray}
and the non-zero ones are; \\
\begin{eqnarray}
F^g(s,t,u;+-++) &=& D(s). \frac{s M_z^2}{2} \sin^2 \theta \nonumber \\
F^g(s,t,u;+-+-) &=& D(s). \frac{s^2}{2} \cos^4 \frac{\theta}{2}\nonumber
\\
F^g(s,t,u:+-+0) &=& D(s). \frac{s^{3/2}M_z}{2\sqrt{2}} (1+
\cos\theta) \sin\theta \nonumber \\
F^g(s,t,u;++-0) &=& D(s). \frac{s^{3/2}M_z}{2\sqrt{2}} (1 -
\cos\theta) \sin\theta \nonumber \\
F^g(s,t,u;++--) &=& D(s). \frac{s M_z^2}{2}
\sin^2\frac{\theta}{2}\nonumber\\
F^g(s,t,u;+--+) &=& D(s). \frac{s^2}{2} \sin^4\frac{\theta}{2}\nonumber \\
F^g(s,t,u;+-0+) &=&
D(s). \frac{s^{3/2}M_z}{2\sqrt{2}}(1-cos\theta) \sin\theta\nonumber \\
F^g(s,t,u;+-0-) &=&
D(s). \frac{s^{3/2}M_z}{2\sqrt{2}}(1+cos\theta) \sin\theta \nonumber \\
F^g(s,t,u;+-00) &=& D(s). \frac{s}{8} ( s+ 4 M_z^2) \sin^2 \theta
\nonumber\\
\end{eqnarray}
with parity symmetry yielding the rest of the amplitudes.  In the
last equation
 the quantity $D(s)$ is
\begin{eqnarray}
D(s) = - \sum_n\frac{\kappa^2}{s-m_n^2} .
\end{eqnarray} 
In the ADD scenario, these resonances almost form a continuum.  The
summation in the last equation can be converted into an integral and has
been explicitly calculated.\cite{five}  The summation however is
UV-divergent and has
to be cut off at some value $M_s$.  $M_s$ would be in order of magnitude
of the Planck scale in $(4 + n)$-dimensions, which then is in the TeV 
range.  Further the effective Newton constant $\kappa$ in 4-dimensions 
would be related to $M_s$ and the compactification size $R$ by:
\begin{eqnarray}\label{eqn}
\kappa^2 R^n = 8 \pi (4\pi)^{n/2} \Gamma (n/2) M_s^{-(n+2)} .
\end{eqnarray}
The result is, assuming there are $n$-extra compact dimensions:
\begin{eqnarray}
- i D(s)= \frac{\kappa^2 s^{n/2 -1} R^n}{\Gamma (n/2)} \left[ \pi + 2 i
I(M_s/\sqrt{s}) \right] ,
\end{eqnarray}
where 
\begin{eqnarray}
I(M_s/ \sqrt{s})= - \sum_{k=1}^{n/2 -1} \frac{1}{2 k}(M_s/\sqrt{s})^{2k} .
\end{eqnarray}
Using equation (\ref{eqn}) , this can be written as:
\begin{eqnarray}
- i D(s) = 8 \pi s^{n/2 -1} M_s^{-n-2} \left[ \pi + 2 i I(
M_s/\sqrt{s}) \right]
\end{eqnarray}
The ADD contribution is thus dependent on a single parameter $M_s$ (
$\approx TeV$) for a given $n$.\\
\indent	In the RS-version of WSQG, the interaction of the discrete set
of resonances of mass $m_n$ can be represented by
\begin{eqnarray}
{\cal L}_I =  -\frac{1}{\Lambda_{\pi}} \sum_n h^{(n)}_{\alpha\beta}
(x) \cdot T^{\alpha\beta}
\end{eqnarray}
where $\Lambda_{\pi}=e^{k r_c \pi} M_{Pl} \approx$ TeV.\cite{six}  The
mass of the 
resonances are given in terms of the zeroes $x_n$ of the Bessel function
$J_1(x)$ by
\begin{eqnarray}
m_n =  k x_n e^{-k r_c \pi} \nonumber \\
\end{eqnarray}
which can be written as
\begin{eqnarray}
m_n =  \frac{k}{M_{Pl}}  m_1  \frac{x_n}{x_1} .
\end{eqnarray}
Phenomenologically, the RS version can be parametrized in terms of two
parameters $(k/M_{pl})$ which is of order 1 and $m_1$, the mass of the
first RS-resonance.\cite{six}  For a given value of $m_1$ and
$(k/M_{Pl})$, the 
summation implied in the evaluation of D(s) cannot be done analytically
but numericallly the
summation converges quite fast.

\section{Results and Discussion}
\hspace{10pt} For the range of parameters indicated therin, our results
are shown in figures 2-9.  There is indeed considerable deviation from the
SM results for the cross-section both in the ADD and the RS version of
WSQG.  The spin-2 nature of the massive gravitons is primarily responsible
for the enhanced contribution of the gravitons in the TeV-range.  This is
responsible for the  $s^2$ factor in their contribution to the amplitudes
and combined with a dimensional coupling this starts dominating in the TeV 
range.  In the ADD scenario, this is essentially because the net
dimensionless parameter governing the amplitude is $(\sqrt{s} / M_s )^6$
for n=2; this is  $O(1)$ in the TeV-range for $\sqrt{s}$ if $M_s$ is in 
the same range. \\
\indent	For the RS scenario, the corresponding factor is $\left(
\frac{k}{M_{Pl}} \right)^2  \frac{s}{m_1^2}$ which once again can be of
$O(1)$ even for small values of the parameter $(k/M_{Pl})$.  The SM
contribution is proportional to $\alpha^2$, so that as  $\sqrt{s}$
increases, one can expect the spin-2 contribution dominating after a
while.  This result of course can be taken seriously only upto a
point.  As $\sqrt{s}/M_s$ or alternatively  $(k/M_{Pl})(\sqrt{s}/m_1)$
becomes larger than one, we can no longer neglect higher order corrections
that renders the amplitude unitary.  Under such circumstances then, the
results numerically are only indicative of considerable deviation from 
SM. \\
\indent	In conclusion, experimental data on the process $\gamma\gamma \to
ZZ$ in the TeV-range and its comparison with well caclulated SM - results
can provide valuable signatures for possible WSQG contributions to the
amplitude.  It is important to note in this connection, that effects
beyond the SM arising out of supersymmetric particle exchange 
contributions are much smaller than the WSQG effects and cannot explain if
large scale departures are obserbved in future experimental data.

\begin{figure}
\includegraphics[angle=270,width=10cm]{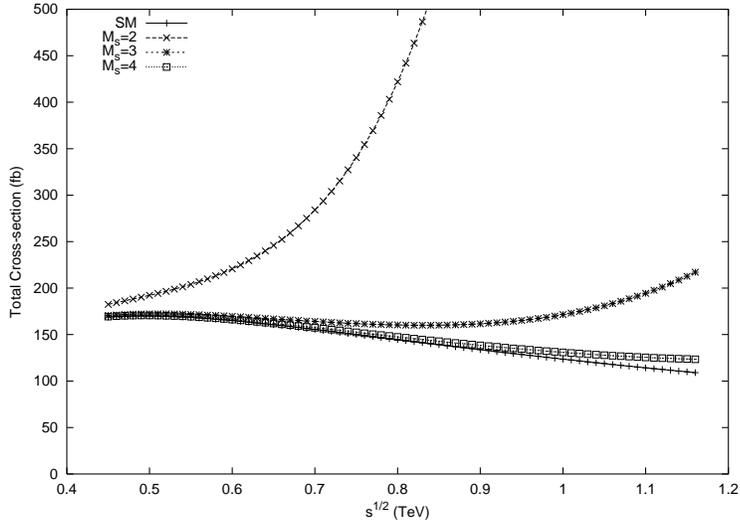}
\caption{The Total cross-section in femto-barnes for the SM contribution
and the SM plus ADD contributions for $M_s = 2, 3, 4$TeV.}
\label{fig2}
\end{figure}
\begin{figure}
\includegraphics[angle=270,width=10cm]{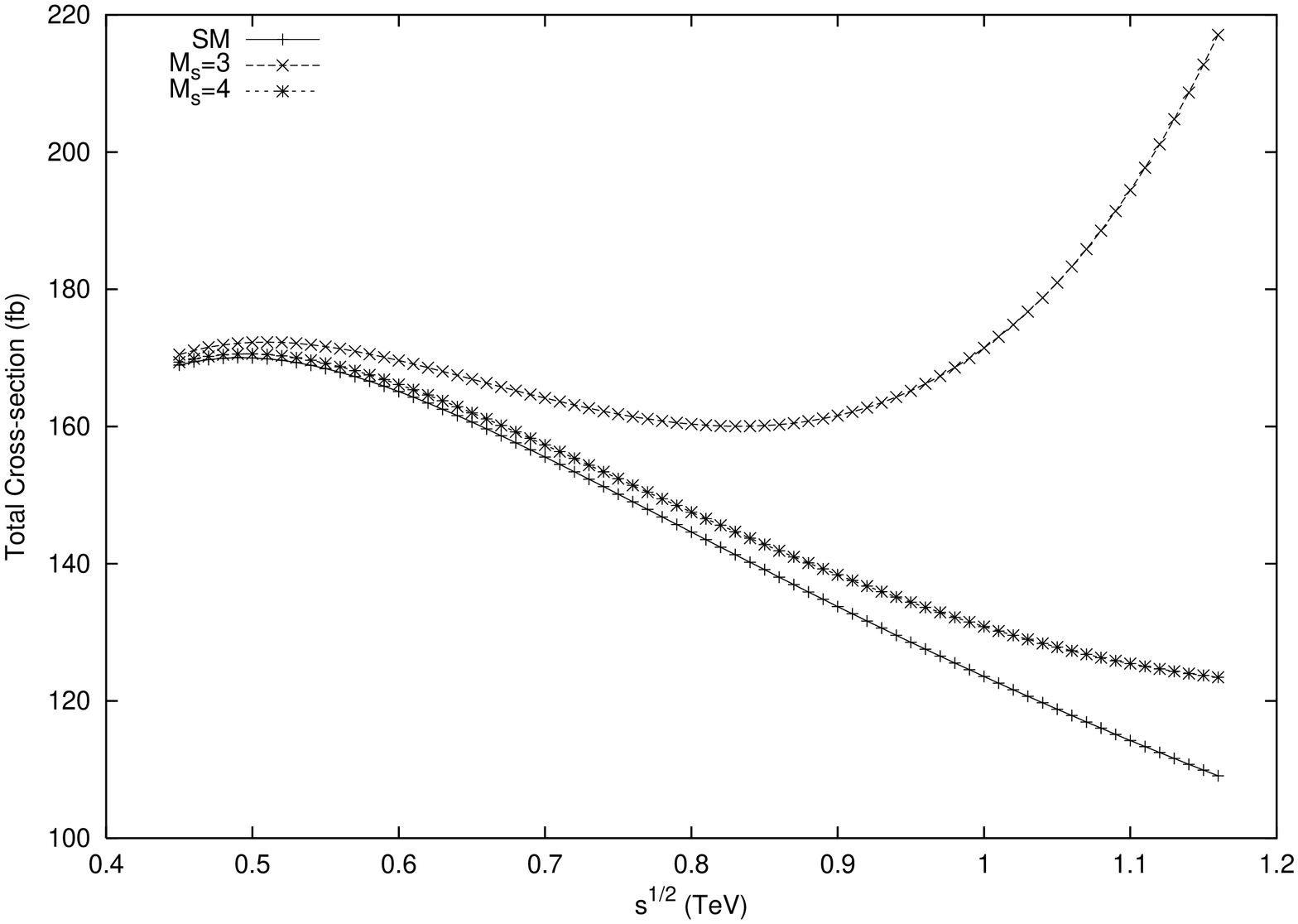}
\caption{The Total cross-section in femto-barnes for the SM contribution
and the SM plus ADD contributions for $M_s = 3, 4$TeV.}
\label{fig4} 
\end{figure}
\begin{figure}
\includegraphics[angle=270,width=10cm]{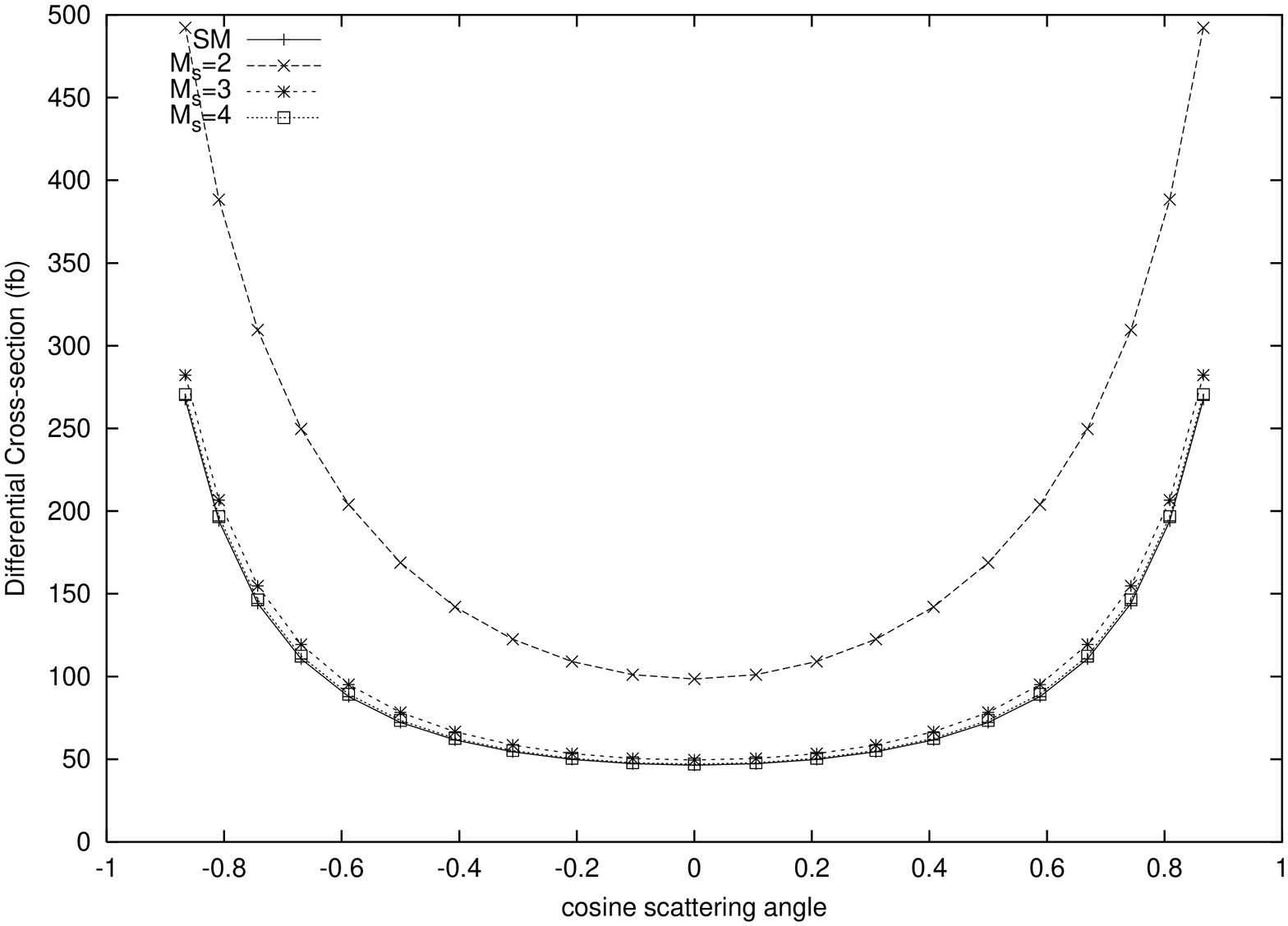}
\caption{The Differential cross-section in femto-barnes for the
$\sqrt{s}=0.8$TeV against $\cos\theta$ for the SM contribution and the SM 
plus ADD contributions for $M_s = 2, 3, 4$TeV.}
\label{fig5}  
\end{figure}
\begin{figure}
\includegraphics[angle=270,width=10cm]{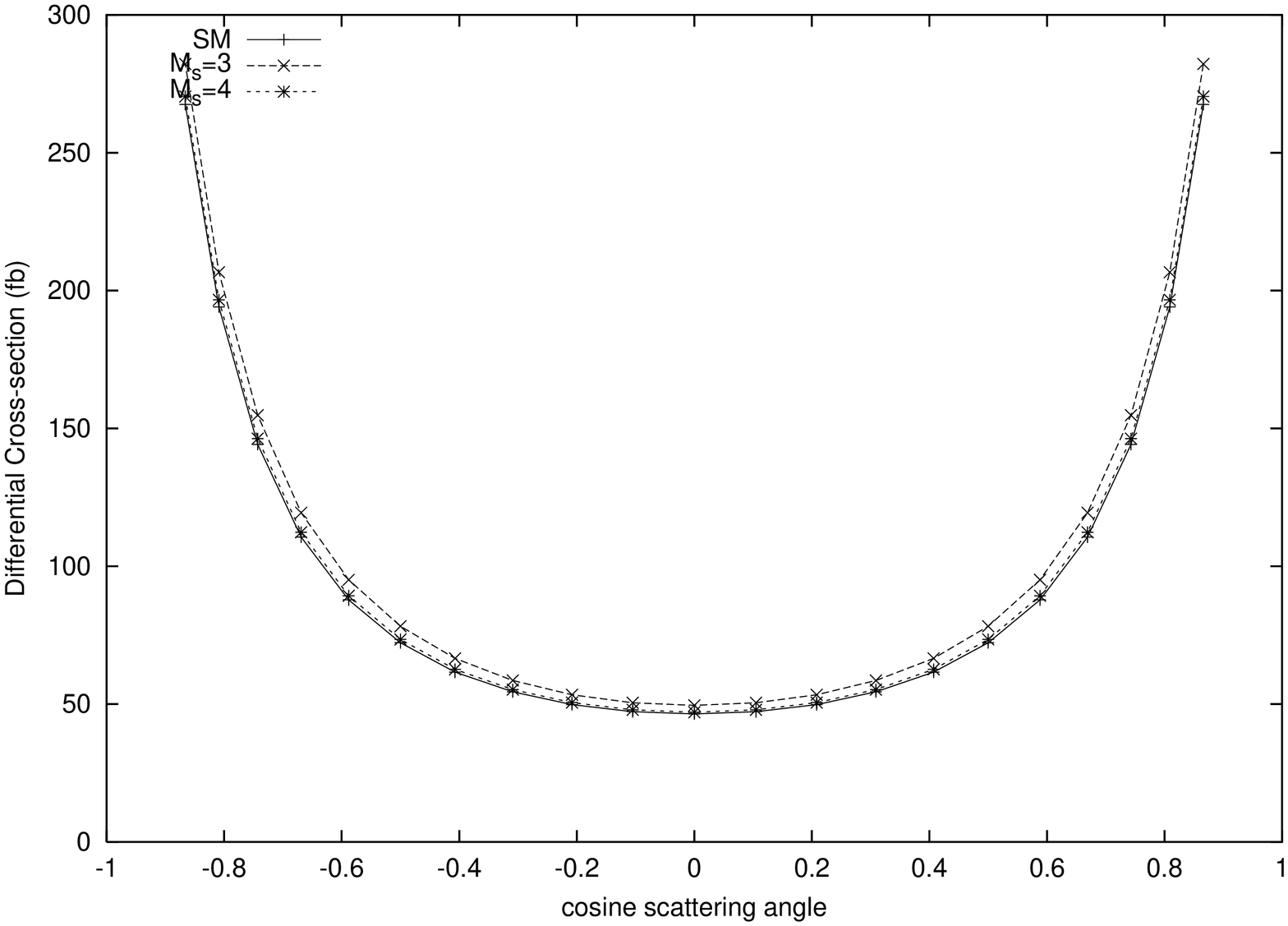}
\caption{The Differential cross-section in femto-barnes for the
$\sqrt{s}=0.8$TeV against $\cos\theta$ for the SM contribution and the SM
plus ADD contributions for $M_s = 3, 4$TeV.}
\label{fig6}  
\end{figure}
\begin{figure}
\includegraphics[angle=270,width=10cm]{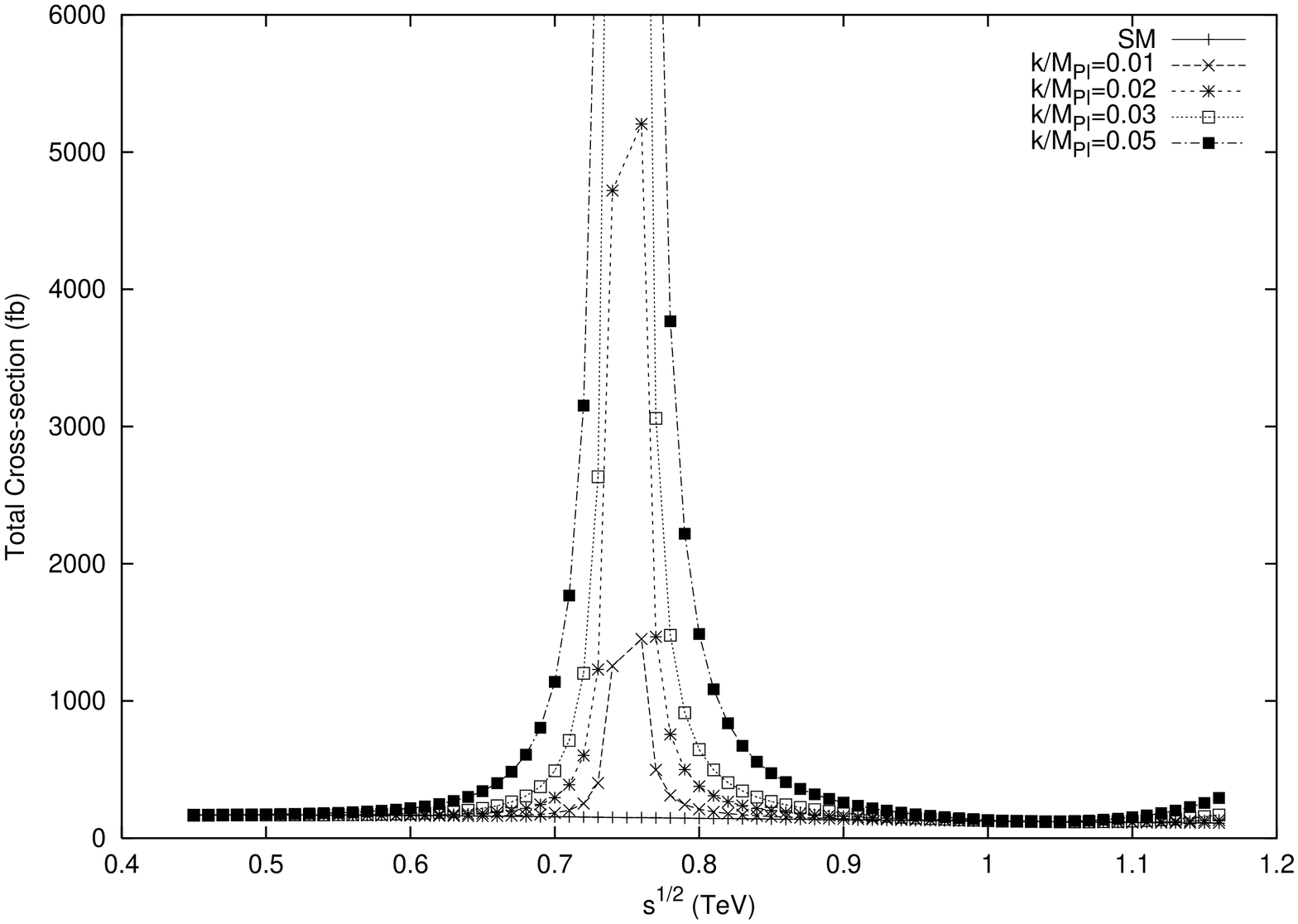}
\caption{The Total cross-section in femto-barnes for the SM contribution
and the SM plus RS contributions for $k/M_{Pl}=0.01, 0.02, 0.03, 0.05$.}
\label{fig7}  
\end{figure}
\begin{figure}
\includegraphics[angle=270,width=10cm]{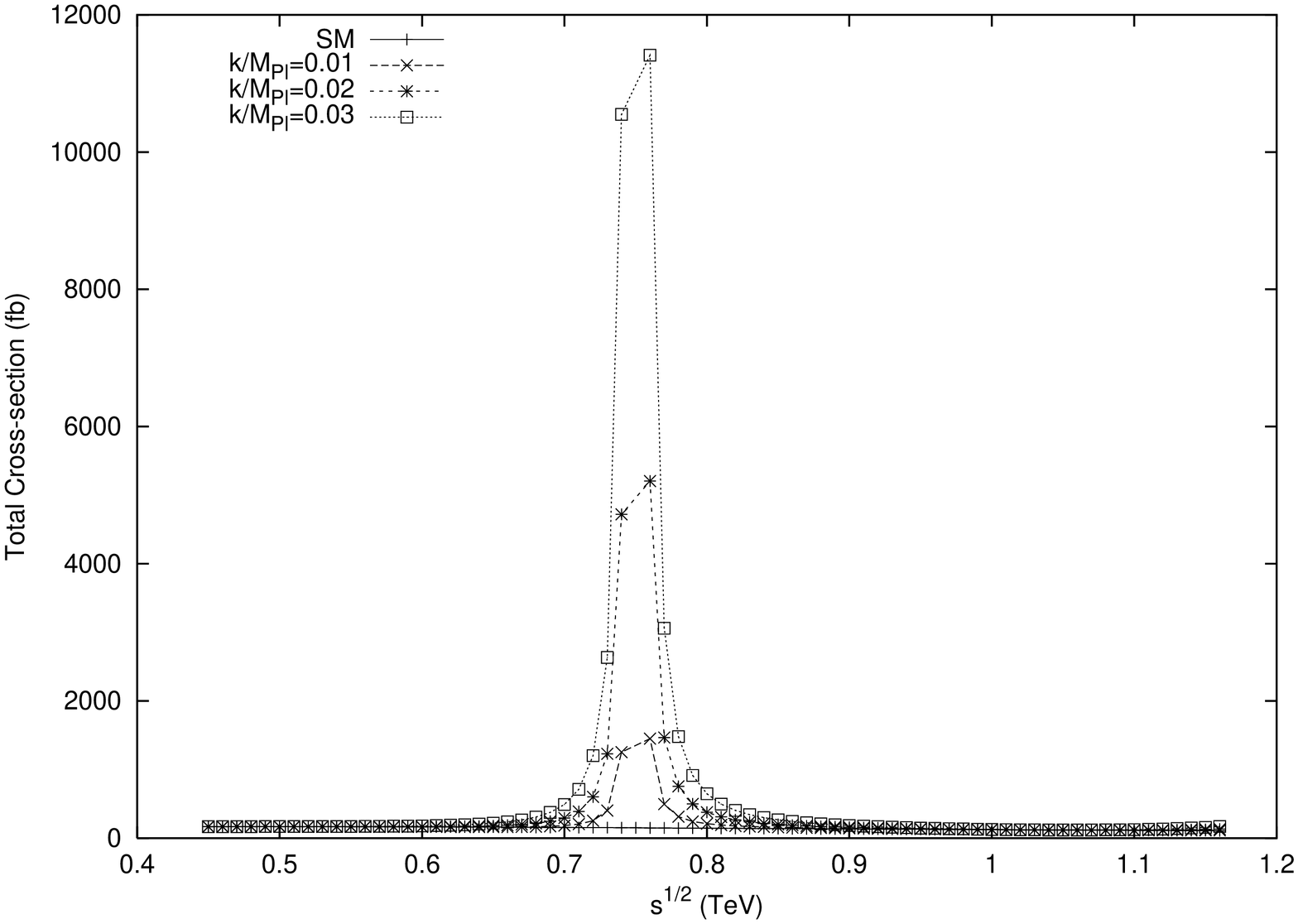}
\caption{The Total cross-section in femto-barnes for the SM contribution
and the SM plus RS contributions for $k/M_{Pl}=0.01, 0.02, 0.03$.}
\label{fig9}
\end{figure}
\begin{figure}
\includegraphics[angle=270,width=10cm]{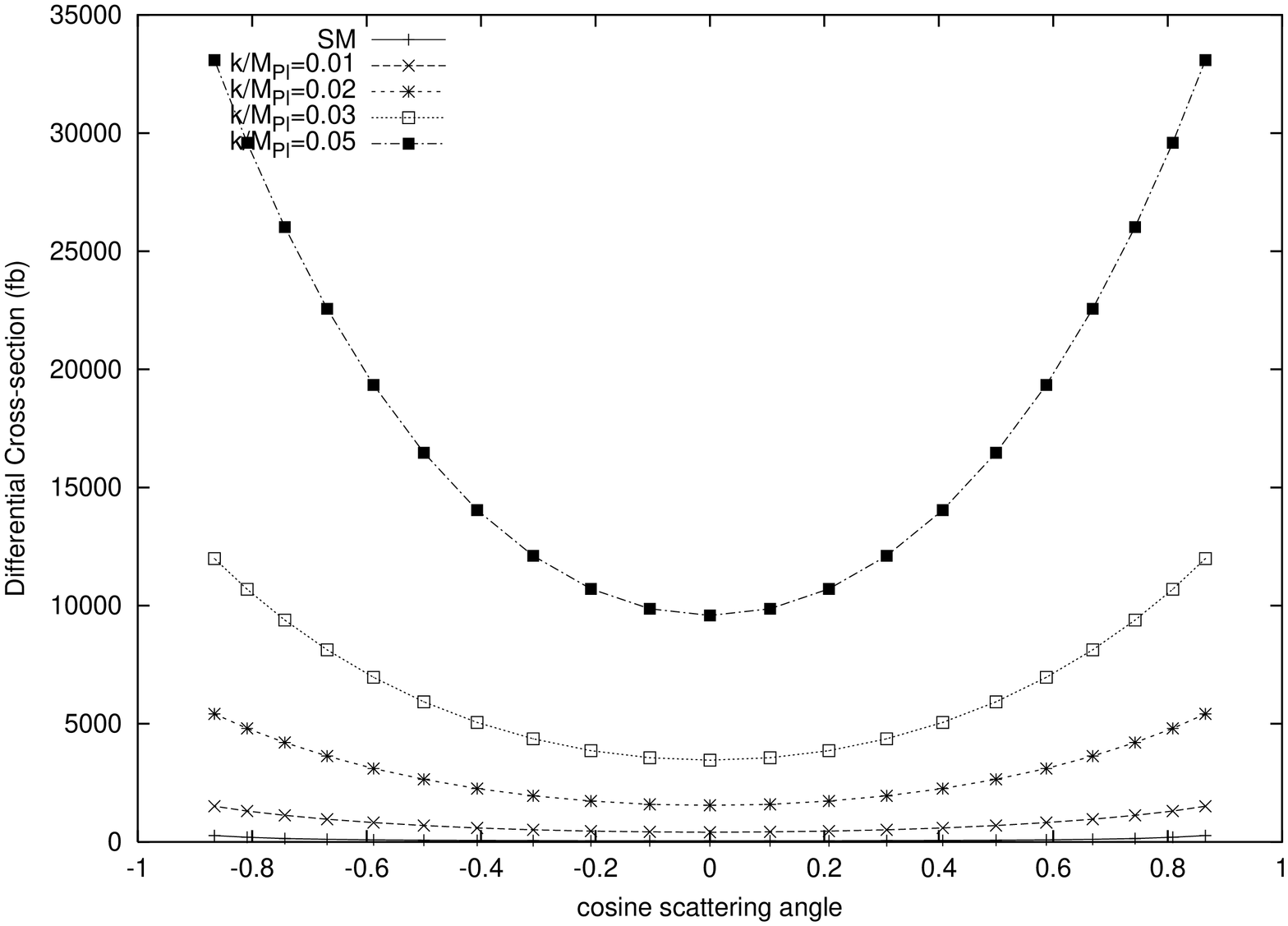}
\caption{The Differential cross-section in femto-barnes for the
$\sqrt{s}=0.8$TeV against $\cos\theta$ for the SM contribution and the SM 
plus RS contributions for $k/M_{Pl}=0.01, 0.02, 0.03, 0.05$.}
\label{fig10} 
\end{figure}
\begin{figure}
\includegraphics[angle=270,width=10cm]{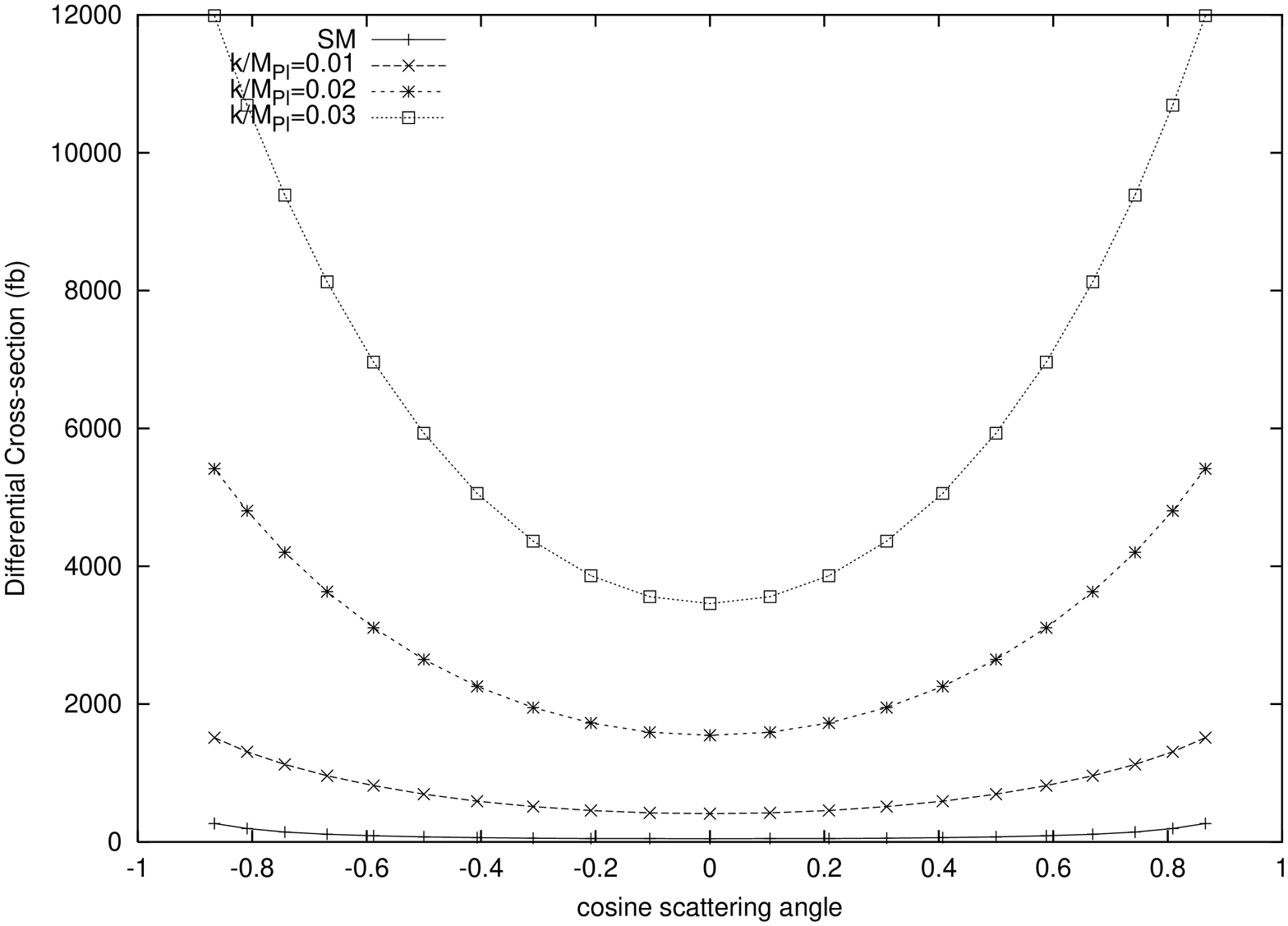}
\caption{The Differential cross-section in femto-barnes for the
$\sqrt{s}=0.8$TeV against $\cos\theta$ for the SM contribution and the SM
plus RS contributions for $k/M_{Pl}=0.01, 0.02, 0.03$.}
\label{fig11}
\end{figure}
\break

\end{document}